# Why Studying Cut-ins? Comparing Cut-ins and Other Lane Changes Based on Naturalistic Driving Data

Yun Lu, *Member, IEEE*, Dejiang Zheng, Rong Su, *Senior Member, IEEE*, Avalpreet Singh Brar, Niels de Boer, and Yong Liang Guan, *Senior Member, IEEE*

*Abstract*—Extensive research has been conducted to explore vehicle lane changes, while the study on cut-ins has not received sufficient attention. The existing studies have not addressed the fundamental question of why studying cut-ins is crucial, despite the extensive investigation into lane changes. To tackle this issue, it is important to demonstrate how cut-ins, as a special type of lane change, differ from other lane changes. In this paper, we explore to compare driving characteristics of cut-ins and other lane changes based on naturalistic driving data. The highD dataset is employed to conduct the comparison. We extract all lane-change events from the dataset and exclude events that are not suitable for our comparison. Lane-change events are then categorized into the cut-in events and other lane-change events based on various gap-based rules. Several performance metrics are designed to measure the driving characteristics of the two types of events. We prove the significant differences between the cut-in behavior and other lane-change behavior by using the Wilcoxon rank-sum test. The results suggest the necessity of conducting specialized studies on cut-ins, offering valuable insights for future research in this field.

## I. INTRODUCTION

A lane change is a driving maneuver where a vehicle moves from one lane to another on a road or highway. This maneuver is one of the most fundamental driving maneuvers and is typically executed to overtake slow-moving vehicles, prepare for upcoming turns, or make way for merging traffic. The lane-change process includes both lateral and longitudinal movements, making the vehicle lane changes closely related to traffic safety and exerting a notable influence on the dynamics of traffic flow [1]. Extensive studies have been conducted to investigate the vehicle lane changes from various perspectives. Some studies focus on inferring the lane-change intention of drivers, which can help reduce the conflicts between the driver and the intelligent vehicle [2]. Some researchers are dedicated to predicting the trajectories of the lane-change vehicles (LCVs), aiming to enhance traffic efficiency, minimize the likelihood of accidents, as well as offer driving strategies for intelligent connected vehicles [3]. Some researchers are committed to mimicking the lane-change maneuver of human drivers for building human-like automated lane-change systems [4].

Vehicle cut-ins refer to a driving maneuver where a lane-change vehicle (LCV) drives into the space ahead of a nearby following vehicle in the target lane, i.e., the target following vehicle (TFV). Cut-in behavior is a special type of lane-change behavior that can frequently occur in traffic. In cut-in scenarios, the TFV may modify its following distance in response to the cut-ins by using emergency braking. The cut-in maneuver has the potential to increase fuel consumption and affect traffic efficiency. It is potentially dangerous and may result in traffic accidents. Besides, it is recognized as a common hazard for the vehicle platoon. It can compromise the integrity of the platoon, making platooning challenging [5].

Recently, several studies have been conducted to analyze and comprehend the cut-in behavior. Wang *et al.* [6] retrieved a large number of cut-in events from naturalistic driving data with the aim of identifying the cut-in characteristics, including motivation, duration, turn signal usage, urgency, and impact. Lu *et al.* [7] built a cut-in control behavior model based on a cognitive architecture to mimic how drivers execute cut-in control when approaching a platoon. In addition, our previous research [8] established a decision model to simulate driver choices regarding whether to proceed with the cut-in and when to execute lane change during the cut-in process. Xiao *et al.* [9] explored how to classify and analyze the driving style of the cut-in behavior.

Besides, some studies have focused on addressing cut-ins by human-driven vehicles for autonomous vehicles (AVs). In [10], a control approach was built based on the driver behavior prediction for AVs to follow the reference trajectory and collaborate with cut-in vehicles. Fu *et al.* [11] built a human-like car-following model for AVs, taking cut-ins in mixed traffic into account. An intelligent speed control strategy was built in [12] to handle uncertain cut-ins. In [13], a method for estimating intentions was integrated into the car-following control system of AVs to identify cut-in vehicles. In [14], a cut-in warning system was built by using multiple rotational camera images. In [15], a cut-in collision warning model was built to handle sudden vehicle cut-ins.

Moreover, substantial efforts have been made for vehicle platoons to manage cut-ins. Basiri *et al.* [16] built a model predictive control method to control a platoon while handling potential cut-in/out. In [17], a sensor fusion algorithm was developed for platoons to identify cut-in vehicles. In [18] and [19], the cooperative adaptive cruise control methods were specifically designed to with the capability of handling vehicle cut-ins. In [20], an optimization algorithm was introduced for connected automated vehicles to optimize trajectories and

This study is supported under the RIE2020 Industry Alignment Fund – Industry Collaboration Projects (IAF-ICP) Funding Initiative, as well as cash and in-kind contribution from the industry partner(s).

Y. Lu, D. Zheng, R. Su, and Y. L. Guan are with the School of Electrical and Electronic Engineering, Nanyang Technological University, Singapore 639798 (e-mail: yun.lu@ntu.edu.sg; dzheng005@e.ntu.edu.sg; rsu@ntu.edu.sg; eylguan@ntu.edu.sg) (The first two authors contributed equally to this work)

N. d. Boer is with the Energy Research Institute, Nanyang Technological University, Singapore 639798 (e-mail: niels.deboer@ntu.edu.sg).

A. S. Brar is with the Continental Automotive Singapore, Singapore 339780 (e-mail: avalpreet.singh.brar@continental-corporation.com)

manage cut-ins, particularly at signalized intersections. Lu *et al.* [21] built a control method for platoons with the aim of minimizing cut-ins while prioritizing road safety.

However, compared to the studies on lane changes, the research on cut-ins has not received sufficient attention. Besides, the existing studies have not addressed the fundamental question of why studying cut-ins is crucial, despite the extensive investigation into lane changes. To tackle this issue, it is significant to demonstrate how cut-ins, as a special type of lane change, differ from other lane changes. Thus, in this paper, we explore to compare the driving characteristics of cut-ins and other lane changes based on naturalistic driving data. Particularly, we employ the highD dataset to conduct the comparison. All lane-change events of the dataset are first extracted. Then, we exclude events that are not suitable for our comparison. Next, lane-change events are categorized into the cut-in events and other lane-change events based on various gap-based rules. After that, several performance metrics are designed to measure the driving characteristics of the two types of events around two key moments in the lane-change process. Finally, the Wilcoxon rank-sum test is applied to evaluate the statistically significant differences between the cut-in behavior and other lane-change behavior.

This paper is organized as follows. Section II presents the description of the cut-in behavior. Section III introduces the analysis method. Section IV describes comparison results. Finally, Section V provides conclusion of this study.

## II. CUT-IN DESCRIPTION

Cut-in behavior is a special type of lane-change behavior. As shown in Figs. 1 (a) and (b), the cut-in process can be divided into two driving phases: cut-in preparation and execution phases. We marked four key instants of the LCV, namely the instants of forming the cut-in intention (i.e., T0), initiating the lateral movement towards the target lane (i.e., T1), crossing the lane mark (i.e., T2), and completing the lateral movement (i.e., T3). Note that since intention refers to the thoughts that one has before the actions, the instant T0 cannot be detected in anyway. During the cut-in preparation phase, the LCV drives within the original lane and adjusts its longitudinal movement to establish an acceptable gap with the TFV. This phase begins at T0 when the LCV generates a cut-in intention and ends at T1 when it initiates the lateral movement toward the target lane. In the cut-in execution phase, the LCV refines its lateral movement to navigate into the target lane. This phase starts at T1 when the lateral movement is initiated and ends at T3 when the lateral movement is completed. Note that the duration of the cut-in preparation phase is significantly influenced by the relative state between the LCV and the TFV at T0. If the relative state at T0 is close to that at T1, the duration of the cut-in preparation phase can be nearly zero.

The cut-in maneuver involves three levels: intention, decision, and control levels. In the intention level, drivers formulate a cut-in intention and decide the target gap according to the surrounding information and vehicle states. In the decision level, drivers decide whether to proceed with the cut-in process and when to execute the lane-change maneuver based on the relative states between the LCV and the TFV. In the control level, drivers conduct the lateral and longitudinal control of the LCV by issuing the steering and acceleration signals, which are transmitted to the vehicle via the steering wheel and throttle/brake, respectively.

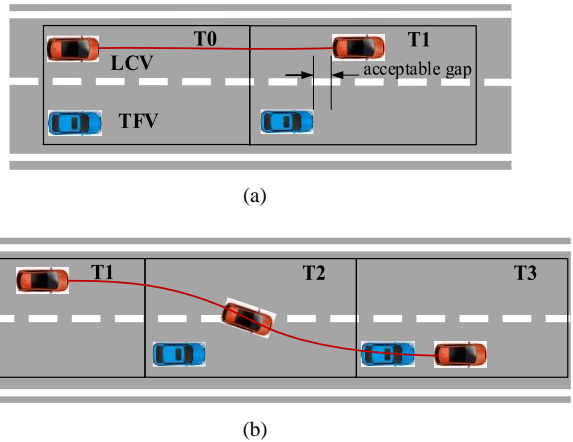

Fig. 1. Cut-in preparation (a) and execution (b) phases.

## III. METHOD

We employed the highD dataset to evaluate the difference between the cut-in behavior and other lane-change behavior. The cut-in events are extracted based on various gap-based rules. Diverse performance metrics are designed to measure the driving characteristics of the two types of behaviors. The Wilcoxon rank-sum test is applied to evaluate the statistically significant differences between the two types of behaviors.

### A. HighD Dataset

The highD dataset contains naturalistic driving data on six different German highways around Cologne in 2017 and 2018 [22]. The driving data was recorded by using a consumer grade drone, whose collection range was about 420 m. The drone was equipped with a 4K-resolution camera. Computer vision algorithms were used to extract more than 45000 km of naturalistic driving behavior from 16.5 h of video recordings. The data included about 11000 vehicles with about 90000 cars and 20000 trucks. For each extracted vehicle, a preprocessed trajectory was available, including positions, velocities, and accelerations. Besides, algorithms annotated the relationships with surrounding vehicles via metrics like time-to-collision and fundamental driving maneuvers such as lane changes for each extracted vehicle. Compared with the extensively used NGSIM, the highD dataset exhibits increased diversity, spanning a wider array of recording sites and featuring extended recording durations. Moreover, the highD dataset presents an increased number of trajectories from truck drivers and an expanded range of mean speed, enabling a more in-depth investigation into the lane-change behavior.

### B. Data Processing

This study focuses on the lane-change maneuvers, which are divided into the cut-in maneuvers and other lane-change maneuvers. We first extract all the lane-change events, with a total count of about 12000. Then, we filter the lane-change events with the following method. First, we exclude the events where the LCV does not have a TFV in lane-change process, as whether a lane-change belongs to a cut-in depends on the gap relationship between the LCV and the TFV. Second, the events where the LCV have different TFVs during the lane-change process are excluded. Third, in the same spirit of [6], the events where the velocity of the LCV or the TFV is no more than 1 m/s are excluded, ensuring that both vehicles are consistently in motion. As a result, a total of 3787 lane-change events are obtained.

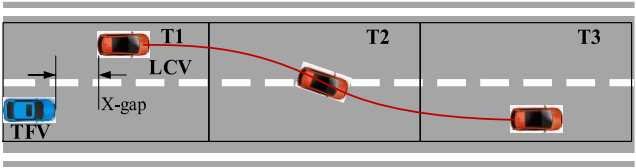

Fig. 2. Lane-change scenario.

Fig. 2 illustrates a general lane-change scenario. In the figure, we highlight three key moments of the LCV that are used and need to be marked at each lane-change event, namely the instants of starting the lateral movement towards the target lane (i.e., T1), crossing the lane mark (i.e., T2), and finishing the lateral movement (i.e., T3). Note that the instant when drivers generate the lane-change intention cannot be observed in anyway in the dataset. Particularly, the moment when the LCV crosses the lane mark is first determined by examining the dataset for the point when the LCV changes its lane ID. Next, the instant when the LCV begins the lateral movement towards the target lane is determined with the following formula:

$$T1 = \min\{t \mid t < T2, v^{(t)} \geq 0, v^{(t+\tau)} \geq v_s, v^{(t_2)} \geq v^{(t_1)} \ (\forall t \leq t_1 \leq t_2 \leq t+\tau_s)\} \quad (1)$$

where $v^{(t)}$ denotes the lateral velocity of the LCV with respect to the target lane at the moment of $t$, $v_s$ is the threshold value for determining the initiation of the lane change, which was taken as 0.15 m/s in the same spirit of [7], and $\tau_s$ represents the observation duration for determining the initiation of the lane change, which was taken to be 1 s. The instant when the LCV finishes the lateral movement is determined by using the following formula:

$$T3 = \min\{t \mid t > T2, |v^{(t_x)}| \leq v_e (\forall t \leq t_x \leq t+\tau_e)\} \quad (2)$$

where $v_e$ denotes the threshold value used to determine the completion of the lane change, which was taken to be 0.1 m/s, and $\tau_e$ is the observation duration for determining the completion of the lane change, which was taken as 1 s.

Next, we extract the cut-in events from the lane-change events. To the best of our knowledge, there is no standardized method for extracting the cut-in events. They can be extracted based on the gap between the LCV and the TFV. Wang et al. [6] identified the lane-change events as the cut-in events when the longitudinal distance between the LCV and the TFV at T1, i.e., X-gap in Fig. 2, is less than 75 m. However, the definition of the cut-in behavior implies that the TFV is near the LCV. Thus, we consider that their defined maximum value of X-gap is too large for the cut-in events. In this paper, we employ five different maximum values of X-gap, i.e., 10, 15, 20, 25, and 30 m, to extract the cut-in events. As shown in Table I, the number of cut-in events obtained with these five different gap limits is 458, 760, 991, 1219, and 1395, respectively. Note that the remaining lane-change events not classified as cut-ins are referred to as "other lane-change events". The number of other lane-change events under the five different gap limits is 3329, 3027, 2796, 2568, and 2392, respectively.

TABLE I
NUMBER OF CUT-IN EVENTS AND OTHER LANE-CHANGE EVENTS UNDER DIFFERENT CUT-IN EXTRACTION GAPS

| Lane change events | X-Gap (m) | | | | |
|---|---|---|---|---|---|
| | 10 | 15 | 20 | 25 | 30 |
| Cut-in | 458 | 760 | 991 | 1219 | 1395 |
| Other lane change | 3329 | 3027 | 2796 | 2568 | 2392 |

C. Evaluation Method

This study aims to compare the driving characteristics of the cut-in events and other lane-change events. The driving characteristics of both the LCV and the TFV are analyzed within certain time intervals around two key instants of the LCV, i.e., the instants of starting the lateral movement towards the target lane (i.e., T1) and crossing the lane mark (i.e., T2). Particularly, we analyzed the driving data at four time intervals around T1, i.e., [T1–4s, T1+1s], [T1–3s, T1+1s] s, [T1–2s, T1+1s], and [T1–1s, T1+1s], and five time intervals around T2, i.e., [T2–2s, T2], [T2–1.5s, T2+0.5s], [T2–1s, T2+1s], [T2–0.5s, T2+1.5s], and [T2, T2+2s].

The driving characteristics were measured by using the maximum acceleration, minimum deceleration, acceleration percentage, velocity change ratio, and cumulative velocity change within a time interval. The acceleration percentage (i.e., $P_a$) within a time interval is computed as

$$P_a = \frac{t_a}{t_T} \quad (3)$$

where $t_T$ is the time interval of interest and $t_a$ denotes the total time in acceleration within the time interval. The velocity change ratio $R_v$ within a time interval is calculated as

$$R_v = \frac{v_{\max} - v_{\min}}{v_{\min}} \quad (4)$$

where $v_{\max}$ and $v_{\min}$ are the maximum and minimum velocities within the time interval of interest, respectively. The cumulative velocity change within a time interval $\Delta v$ is computed as

$$\Delta v = \sum_{j=0}^{n-1} |a_j \cdot \Delta t| \quad (5)$$

where $n$ is the total step number of the time interval of interest, $a_j$ denotes the acceleration at time step $j$, and $\Delta t$ is the duration of a time step, which equals to 0.04 s in the original highD dataset.

Furthermore, we conduct a statistical analysis to assess the significance of the difference in the driving characteristics between the cut-in events and other lane-change events. The statistical analysis is conducted with the Wilcoxon rank-sum test at the 95% confidence interval, which is a non-parametric method suited for comparing distributions when assumptions of normality are not met.

## IV. Results

### A. Driving Characteristics of LCV around T1

Table II compares the cumulative velocity change of the LCV between the cut-in events and other lane-change events within various time intervals around T1 under different cut-in extraction gaps. We see that all the mean cumulative velocity changes of the LCV under the cut-in events are larger than those under other lane-change events. According to the Wilcoxon rank-sum test, statistically significant differences are observed in each comparison, with all p-values being close to zero.

Table III compares the velocity change ratio of the LCV between the cut-in events and other lane-change events within various time intervals around T1 across different cut-in extraction gaps. We see that all the mean velocity change ratios of the LCV under the cut-in events are greater than those under other lane-change events. Statistically significant differences are identified in each comparison, with all p-values approaching zero.

The reason behind these results could be that the LCV in the cut-in events expends more control effort to generate a safe gap between the TFV before initiating the lane change.

TABLE II
COMPARISON OF CUMULATIVE VELOCITY CHANGE OF THE LCV BETWEEN CUT-IN EVENTS AND OTHER LANE-CHANGE EVENTS WITHIN VARIOUS TIME INTERVALS AROUND T1 UNDER DIFFERENT CUT-IN EXTRACTION GAPS [MEAN AND STANDARD DEVIATION (STD)]

| X-Gap (m) | Time interval (s) | Cut-in | Other lane-change | p-value |
|---|---|---|---|---|
| 10 | [T1–4,T1+1] | 0.918 (0.471) | 0.873 (0.572) | 0.005 |
| | [T1–3,T1+1] | 0.799 (0.382) | 0.760 (0.474) | 0.004 |
| | [T1–2,T1+1] | 0.669 (0.292) | 0.614 (0.370) | <0.001 |
| | [T1–1,T1+1] | 0.476 (0.205) | 0.454 (0.275) | <0.001 |
| 15 | [T1–4,T1+1] | 0.948 (0.469) | 0.864 (0.582) | <0.001 |
| | [T1–3,T1+1] | 0.828 (0.396) | 0.752 (0.480) | <0.001 |
| | [T1–2,T1+1] | 0.672 (0.304) | 0.610 (0.370) | <0.001 |
| | [T1–1,T1+1] | 0.478 (0.229) | 0.454 (0.276) | <0.001 |
| 20 | [T1–4,T1+1] | 0.966 (0.473) | 0.847 (0.586) | <0.001 |
| | [T1–3,T1+1] | 0.854 (0.396) | 0.736 (0.483) | <0.001 |
| | [T1–2,T1+1] | 0.697 (0.315) | 0.600 (0.370) | <0.001 |
| | [T1–1,T1+1] | 0.497 (0.238) | 0.449 (0.276) | <0.001 |
| 25 | [T1–4,T1+1] | 0.989 (0.481) | 0.819 (0.591) | <0.001 |
| | [T1–3,T1+1] | 0.874 (0.409) | 0.719 (0.486) | <0.001 |
| | [T1–2,T1+1] | 0.719 (0.333) | 0.583 (0.365) | <0.001 |
| | [T1–1,T1+1] | 0.506 (0.246) | 0.443 (0.272) | <0.001 |
| 30 | [T1–4,T1+1] | 1.020 (0.500) | 0.791 (0.585) | <0.001 |
| | [T1–3,T1+1] | 0.897 (0.420) | 0.706 (0.485) | <0.001 |
| | [T1–2,T1+1] | 0.728 (0.338) | 0.569 (0.365) | <0.001 |
| | [T1–1,T1+1] | 0.513 (0.248) | 0.434 (0.267) | <0.001 |

TABLE III
COMPARISON OF VELOCITY CHANGE RATIO OF THE LCV BETWEEN CUT-IN EVENTS AND OTHER LANE-CHANGE EVENTS WITHIN VARIOUS TIME INTERVALS AROUND T1 UNDER DIFFERENT CUT-IN EXTRACTION GAPS [MEAN AND (STD)]

| X-Gap (m) | Time interval (s) | Cut-in | Other lane-change | p-value |
|---|---|---|---|---|
| 10 | [T1–4,T1+1] | 0.027 (0.015) | 0.025 (0.019) | <0.001 |
| | [T1–3,T1+1] | 0.024 (0.012) | 0.022 (0.015) | <0.001 |
| | [T1–2,T1+1] | 0.020 (0.009) | 0.018 (0.012) | <0.001 |
| | [T1–1,T1+1] | 0.014 (0.006) | 0.014 (0.009) | 0.002 |
| 15 | [T1–4,T1+1] | 0.028 (0.015) | 0.024 (0.019) | <0.001 |
| | [T1–3,T1+1] | 0.024 (0.013) | 0.021 (0.016) | <0.001 |
| | [T1–2,T1+1] | 0.020 (0.010) | 0.018 (0.012) | <0.001 |
| | [T1–1,T1+1] | 0.014 (0.007) | 0.014 (0.009) | <0.001 |
| 20 | [T1–4,T1+1] | 0.029 (0.016) | 0.024 (0.019) | <0.001 |
| | [T1–3,T1+1] | 0.026 (0.014) | 0.021 (0.015) | <0.001 |
| | [T1–2,T1+1] | 0.021 (0.010) | 0.018 (0.012) | <0.001 |
| | [T1–1,T1+1] | 0.015 (0.007) | 0.014 (0.009) | <0.001 |
| 25 | [T1–4,T1+1] | 0.030 (0.017) | 0.023 (0.019) | <0.001 |
| | [T1–3,T1+1] | 0.026 (0.014) | 0.020 (0.015) | <0.001 |
| | [T1–2,T1+1] | 0.021 (0.011) | 0.017 (0.012) | <0.001 |
| | [T1–1,T1+1] | 0.015 (0.008) | 0.013 (0.009) | <0.001 |
| 30 | [T1–4,T1+1] | 0.031 (0.018) | 0.022 (0.018) | <0.001 |
| | [T1–3,T1+1] | 0.027 (0.014) | 0.020 (0.015) | <0.001 |
| | [T1–2,T1+1] | 0.022 (0.011) | 0.017 (0.012) | <0.001 |
| | [T1–1,T1+1] | 0.016 (0.008) | 0.013 (0.009) | <0.001 |

TABLE IV
COMPARISON OF ACCELERATION PERCENTAGE OF THE LCV BETWEEN CUT-IN EVENTS AND OTHER LANE-CHANGE EVENTS WITHIN VARIOUS TIME INTERVALS AROUND T2 UNDER DIFFERENT CUT-IN EXTRACTION GAPS [MEAN AND (STD)]

| X-Gap (m) | Time interval (s) | Cut-in | Other lane-change | p-value |
|---|---|---|---|---|
| 10 | [T2–2,T2] | 0.431 (0.465) | 0.559 (0.467) | <0.001 |
| | [T2–1.5, T2+0.5] | 0.400 (0.455) | 0.551 (0.471) | <0.001 |
| | [T2–1, T2+1] | 0.374 (0.447) | 0.543 (0.474) | <0.001 |
| | [T2–0.5, T2+1.5] | 0.342 (0.442) | 0.513 (0.476) | <0.001 |
| | [T2, T2+2] | 0.331 (0.438) | 0.489 (0.466) | <0.001 |
| 15 | [T2–2,T2] | 0.479 (0.468) | 0.559 (0.468) | <0.001 |
| | [T2–1.5, T2+0.5] | 0.459 (0.465) | 0.552 (0.472) | <0.001 |
| | [T2–1, T2+1] | 0.440 (0.463) | 0.543 (0.475) | <0.001 |
| | [T2–0.5, T2+1.5] | 0.406 (0.461) | 0.514 (0.476) | <0.001 |
| | [T2, T2+2] | 0.382 (0.451) | 0.492 (0.466) | <0.001 |
| 20 | [T2–2,T2] | 0.498 (0.469) | 0.559 (0.468) | <0.001 |
| | [T2–1.5, T2+0.5] | 0.477 (0.466) | 0.553 (0.472) | <0.001 |
| | [T2–1, T2+1] | 0.458 (0.465) | 0.545 (0.476) | <0.001 |
| | [T2–0.5, T2+1.5] | 0.422 (0.464) | 0.518 (0.477) | <0.001 |
| | [T2, T2+2] | 0.401 (0.455) | 0.494 (0.467) | <0.001 |
| 25 | [T2–2,T2] | 0.507 (0.468) | 0.560 (0.468) | 0.001 |
| | [T2–1.5, T2+0.5] | 0.484 (0.467) | 0.556 (0.472) | <0.001 |
| | [T2–1, T2+1] | 0.467 (0.466) | 0.548 (0.476) | <0.001 |
| | [T2–0.5, T2+1.5] | 0.432 (0.466) | 0.521 (0.477) | <0.001 |
| | [T2, T2+2] | 0.412 (0.459) | 0.497 (0.466) | <0.001 |
| 30 | [T2–2,T2] | 0.518 (0.469) | 0.558 (0.468) | 0.013 |
| | [T2–1.5, T2+0.5] | 0.497 (0.469) | 0.554 (0.472) | <0.001 |
| | [T2–1, T2+1] | 0.480 (0.469) | 0.547 (0.476) | <0.001 |
| | [T2–0.5, T2+1.5] | 0.447 (0.468) | 0.519 (0.478) | <0.001 |
| | [T2, T2+2] | 0.426 (0.460) | 0.495 (0.467) | <0.001 |

TABLE V
COMPARISON OF MAXIMUM ACCELERATION OF THE LCV BETWEEN CUT-IN EVENTS AND OTHER LANE-CHANGE EVENTS WITHIN VARIOUS TIME INTERVALS AROUND T2 UNDER DIFFERENT CUT-IN EXTRACTION GAPS [MEAN AND (STD)]

| X-Gap (m) | Time interval (s) | Lane-change events Cut-in | Lane-change events Other lane-change | p-value |
|---|---|---|---|---|
| 10 | [T2–2,T2] | 0.199 (0.122) | 0.294 (0.192) | <0.001 |
| 10 | [T2–1.5, T2+0.5] | 0.194 (0.124) | 0.288 (0.190) | <0.001 |
| 10 | [T2–1, T2+1] | 0.198 (0.127) | 0.291 (0.194) | <0.001 |
| 10 | [T2–0.5, T2+1.5] | 0.199 (0.129) | 0.278 (0.184) | <0.001 |
| 10 | [T2, T2+2] | 0.192 (0.128) | 0.255 (0.169) | <0.001 |
| 15 | [T2–2,T2] | 0.221 (0.145) | 0.301 (0.196) | <0.001 |
| 15 | [T2–1.5, T2+0.5] | 0.212 (0.142) | 0.298 (0.196) | <0.001 |
| 15 | [T2–1, T2+1] | 0.210 (0.143) | 0.294 (0.194) | <0.001 |
| 15 | [T2–0.5, T2+1.5] | 0.209 (0.138) | 0.280 (0.183) | <0.001 |
| 15 | [T2, T2+2] | 0.197 (0.131) | 0.256 (0.169) | <0.001 |
| 20 | [T2–2,T2] | 0.241 (0.158) | 0.302 (0.198) | <0.001 |
| 20 | [T2–1.5, T2+0.5] | 0.231 (0.154) | 0.301 (0.198) | <0.001 |
| 20 | [T2–1, T2+1] | 0.217 (0.145) | 0.303 (0.200) | <0.001 |
| 20 | [T2–0.5, T2+1.5] | 0.215 (0.144) | 0.295 (0.195) | <0.001 |
| 20 | [T2, T2+2] | 0.205 (0.137) | 0.265 (0.176) | <0.001 |
| 25 | [T2–2,T2] | 0.244 (0.158) | 0.304 (0.199) | <0.001 |
| 25 | [T2–1.5, T2+0.5] | 0.235 (0.156) | 0.307 (0.202) | <0.001 |
| 25 | [T2–1, T2+1] | 0.229 (0.150) | 0.304 (0.201) | <0.001 |
| 25 | [T2–0.5, T2+1.5] | 0.222 (0.145) | 0.296 (0.197) | <0.001 |
| 25 | [T2, T2+2] | 0.213 (0.142) | 0.266 (0.177) | <0.001 |
| 30 | [T2–2,T2] | 0.258 (0.169) | 0.306 (0.202) | <0.001 |
| 30 | [T2–1.5, T2+0.5] | 0.247 (0.165) | 0.304 (0.200) | <0.001 |
| 30 | [T2–1, T2+1] | 0.241 (0.159) | 0.306 (0.204) | <0.001 |
| 30 | [T2–0.5, T2+1.5] | 0.232 (0.152) | 0.297 (0.197) | <0.001 |
| 30 | [T2, T2+2] | 0.214 (0.144) | 0.267 (0.176) | <0.001 |

### B. Driving Characteristics of LCV around T2

Table IV compares the acceleration percentage of the LCV between the cut-in events and other lane-change events within various time intervals around T2 under different cut-in extraction gaps. We see that all the mean acceleration percentages of the LCV under the cut-in events are smaller than those under other lane-change events. Moreover, each comparison reveals statistically significant differences, with all p-values approaching zero.

Table V compares the maximum acceleration of the LCV between the cut-in events and other lane-change events within various time intervals around T2 under different cut-in extraction gaps. We see that all the average maximum accelerations of the LCV under cut-in events are smaller than those under other lane-change events. Statistically significant differences are observed in each comparison.

The reason behind these results could be that the target gaps in the cut-in events are smaller than those in other lane-change events, thereby restricting the acceleration of the LCV during the execution of lane change.

### C. Driving Characteristics of TFV around T1

Table VI compares the minimum deceleration of the TFV between the cut-in events and other lane-change events within various time intervals around T1 under different cut-in extraction gaps. We see that all the average minimum decelerations of the TFV under the cut-in events are smaller than those under other lane-change events. Each comparison presents statistically significant differences.

The reason behind these results could be that the TFV in cut-in events tends to apply more urgent decelerations than the TFV in other lane-change events.

TABLE VI
COMPARISON OF MINIMUM DECELERATION OF THE TFV BETWEEN CUT-IN EVENTS AND OTHER LANE-CHANGE EVENTS WITHIN VARIOUS TIME INTERVALS AROUND T1 UNDER DIFFERENT CUT-IN EXTRACTION GAPS [MEAN AND (STD)]

| X-Gap (m) | Time interval (s) | Lane-change events Cut-in | Lane-change events Other lane-change | p-value |
|---|---|---|---|---|
| 10 | [T1–4,T1+1] | -0.120 (0.055) | -0.086 (0.114) | <0.001 |
| 10 | [T1–3,T1+1] | -0.117 (0.054) | -0.083 (0.112) | <0.001 |
| 10 | [T1–2,T1+1] | -0.115 (0.052) | -0.082 (0.112) | <0.001 |
| 10 | [T1–1,T1+1] | -0.105 (0.050) | -0.077 (0.108) | <0.001 |
| 15 | [T1–4,T1+1] | -0.111 (0.057) | -0.093 (0.117) | <0.001 |
| 15 | [T1–3,T1+1] | -0.112 (0.056) | -0.090 (0.115) | <0.001 |
| 15 | [T1–2,T1+1] | -0.110 (0.054) | -0.089 (0.115) | <0.001 |
| 15 | [T1–1,T1+1] | -0.101 (0.050) | -0.085 (0.113) | 0.002 |
| 20 | [T1–4,T1+1] | -0.108 (0.058) | -0.098 (0.118) | 0.015 |
| 20 | [T1–3,T1+1] | -0.109 (0.058) | -0.095 (0.117) | 0.003 |
| 20 | [T1–2,T1+1] | -0.108 (0.055) | -0.092 (0.115) | 0.002 |
| 20 | [T1–1,T1+1] | -0.101 (0.052) | -0.087 (0.113) | 0.005 |
| 25 | [T1–4,T1+1] | -0.114 (0.063) | -0.100 (0.118) | 0.002 |
| 25 | [T1–3,T1+1] | -0.112 (0.060) | -0.097 (0.116) | 0.001 |
| 25 | [T1–2,T1+1] | -0.110 (0.057) | -0.096 (0.117) | 0.002 |
| 25 | [T1–1,T1+1] | -0.104 (0.055) | -0.089 (0.112) | 0.002 |
| 30 | [T1–4,T1+1] | -0.114 (0.063) | -0.101 (0.118) | 0.003 |
| 30 | [T1–3,T1+1] | -0.115 (0.063) | -0.098 (0.116) | <0.001 |
| 30 | [T1–2,T1+1] | -0.112 (0.058) | -0.097 (0.116) | 0.002 |
| 30 | [T1–1,T1+1] | -0.107 (0.057) | -0.090 (0.111) | <0.001 |

### D. Driving Characteristics of TFV around T2

Table VII compares the cumulative velocity change of the TFV between the cut-in events and other lane-change events within various time intervals around T2 under different cut-in extraction gaps. We see that all the mean cumulative velocity changes of the LCV under the cut-in events are smaller than those under other lane-change events. Statistically significant differences are observed in each comparison.

The reason behind these results could be that the TFV in the cut-in events tends to maintain a more stable velocity to address the cut-ins compared to the TFV in other lane change events.

### V. CONCLUSION

This paper has investigated to compare and analyze the driving characteristics of cut-ins and other lane changes based on the highD dataset. We extracted all lane-change events from the dataset and excluded events that were not suitable for the comparison. Lane-change events were divided into the cut-in events and other lane-change events by using various gap-based rules. Several performance metrics were designed to measure the driving characteristics of the two types of events around two key moments in the lane-change process. We demonstrated the significant differences between the cut-in behavior and other lane-change behavior with the

Wilcoxon rank-sum test. The results suggest the necessity of conducting a specialized study on cut-ins, offering valuable insights for future research in this field.

TABLE VII
COMPARISON OF CUMULATIVE VELOCITY CHANGE OF THE TFV BETWEEN CUT-IN EVENTS AND OTHER LANE-CHANGE EVENTS WITHIN VARIOUS TIME INTERVALS AROUND T2 UNDER DIFFERENT CUT-IN EXTRACTION GAPS [MEAN AND (STD)]

| X-Gap (m) | Time interval (s) | Lane-change events | | p-value |
|---|---|---|---|---|
| | | Cut-in | Other lane-change | |
| 10 | [T2–2, T2] | 0.267 (0.128) | 0.332 (0.182) | <0.001 |
| | [T2–1.5, T2+0.5] | 0.265 (0.119) | 0.349 (0.198) | <0.001 |
| | [T2–1, T2+1] | 0.270 (0.134) | 0.369 (0.219) | <0.001 |
| | [T2–0.5, T2+1.5] | 0.276 (0.146) | 0.384 (0.230) | <0.001 |
| | [T2, T2+2] | 0.292 (0.155) | 0.397 (0.240) | <0.001 |
| 15 | [T2–2, T2] | 0.266 (0.126) | 0.341 (0.188) | <0.001 |
| | [T2–1.5, T2+0.5] | 0.265 (0.120) | 0.358 (0.201) | <0.001 |
| | [T2–1, T2+1] | 0.272 (0.138) | 0.380 (0.224) | <0.001 |
| | [T2–0.5, T2+1.5] | 0.291 (0.156) | 0.392 (0.233) | <0.001 |
| | [T2, T2+2] | 0.298 (0.160) | 0.406 (0.245) | <0.001 |
| 20 | [T2–2, T2] | 0.270 (0.124) | 0.349 (0.195) | <0.001 |
| | [T2–1.5, T2+0.5] | 0.270 (0.120) | 0.364 (0.209) | <0.001 |
| | [T2–1, T2+1] | 0.279 (0.137) | 0.384 (0.230) | <0.001 |
| | [T2–0.5, T2+1.5] | 0.298 (0.153) | 0.397 (0.239) | <0.001 |
| | [T2, T2+2] | 0.308 (0.157) | 0.414 (0.253) | <0.001 |
| 25 | [T2–2, T2] | 0.278 (0.133) | 0.350 (0.196) | <0.001 |
| | [T2–1.5, T2+0.5] | 0.284 (0.138) | 0.366 (0.210) | <0.001 |
| | [T2–1, T2+1] | 0.294 (0.150) | 0.386 (0.230) | <0.001 |
| | [T2–0.5, T2+1.5] | 0.310 (0.164) | 0.398 (0.241) | <0.001 |
| | [T2, T2+2] | 0.329 (0.179) | 0.414 (0.253) | <0.001 |
| 30 | [T2–2, T2] | 0.289 (0.141) | 0.348 (0.198) | <0.001 |
| | [T2–1.5, T2+0.5] | 0.296 (0.151) | 0.363 (0.209) | <0.001 |
| | [T2–1, T2+1] | 0.305 (0.162) | 0.384 (0.229) | <0.001 |
| | [T2–0.5, T2+1.5] | 0.326 (0.186) | 0.395 (0.237) | <0.001 |
| | [T2, T2+2] | 0.339 (0.196) | 0.411 (0.247) | <0.001 |